\documentclass[useAMS,usenatbib]{mn2e}

\usepackage{epsfig}
\usepackage{amsmath}
\usepackage{amssymb}
\usepackage{natbib}
\usepackage{threeparttable} 
\usepackage{multirow}

\usepackage{epstopdf}

\newcommand{\chan}{\textit{Chandra}}

\newcommand{\Msun}{\mathrm{M}_{\odot}}
\newcommand{\lum}{\mathrm{erg~s}^{-1}}
\newcommand{\flux}{\mathrm{erg~cm}^{-2}~\mathrm{s}^{-1}}

\newcommand{\nh}{\mathrm{cm^{-2}}}

\newcommand{\cxo}{CXOGClb J174804.8--244648}
\newcommand{\igr}{IGR J17480--2446}
\newcommand{\exo}{EXO~0748--676}
\newcommand{\ks}{KS~1731--260}
\newcommand{\mxb}{MXB~1659--29}
\newcommand{\xte}{XTE~J1701--462}

\def \mnras {MNRAS}
\def \apj {ApJ}

\def \apjl {ApJL}
\def \aap {A\&A}

\def \prc {PhRvC}

\def \pre {PhRvE}

\title[Crust cooling of the Terzan 5 X-ray pulsar]{Evidence for crust cooling in the transiently accreting 11-Hz X-ray pulsar in the globular cluster Terzan 5}
\author[N. Degenaar E. F. Brown \& R. Wijnands]
{N. Degenaar$^{1}$\thanks{e-mail: degenaar@uva.nl}, E. F. Brown$^{2}$ \& R. Wijnands$^{1}$\\
$^{1}$Astronomical Institute "Anton Pannekoek", 
University of Amsterdam, 
Postbus 94249, 1090 GE Amsterdam, the Netherlands\\
$^{2}$Department of Physics and Astronomy, 
Michigan State University, 
East Lansing, MI 48824, USA
\vspace{-0.3cm}
}

\begin{document}

\date{Accepted 2011 September 28. Received 2011 August 30; in original form 2011 July 26}

\pagerange{\pageref{firstpage}--\pageref{lastpage}} \pubyear{0000}

\maketitle

\label{firstpage}

\begin{abstract} 
The temporal heating and subsequent cooling of the crusts of transiently accreting neutron stars carries unique information about their structure and a variety of nuclear reaction processes. We report on a new \chan\ Director's Discretionary Time observation of the globular cluster Terzan 5, aimed to monitor the transiently accreting 11-Hz X-ray pulsar \igr\ after the cessation of its recent 10-week long accretion outburst. During the observation, which was performed $\simeq125$ days into quiescence, the source displays a thermal spectrum that fits to a neutron star atmosphere model with a temperature for an observer at infinity of $kT^{\infty} \simeq 92$~eV. This is $\simeq10\%$ lower than found $\simeq75$ days earlier, yet $\simeq 20 \%$ higher than the quiescent base level measured prior to the recent outburst. This can be interpreted as cooling of the accretion-heated neutron star crust, and implies that crust cooling is observable after short accretion episodes. Comparison with neutron star thermal evolution simulations indicates that substantial heat must be released at shallow depth inside the neutron star, which is not accounted for in current nuclear heating models. 
\end{abstract}


\section{Introduction}\label{sec:intro}
Neutron stars are the densest directly observable stellar objects in our Universe and constitute ideal astrophysical laboratories to study matter under extreme physical conditions. The outer layer of a neutron star, its crust, covers about one tenth of the total stellar radius and consists of ions, electrons and neutrons. The structure and composition of the crust play an important role in the emission of gravitational waves and the evolution of the neutron star's magnetic field \citep{brown1998_2,ushomirsky2000,horowitz2009}. Furthermore, studying the crusts of neutron stars allows the investigation of a variety of nuclear reaction processes \citep{haensel2008,horowitz2008}.

When residing in a low-mass X-ray binary (LMXB), a neutron star strips off the outer gaseous layers of a stellar companion through Roche-lobe overflow, and this matter is accreted onto the neutron star. Transient systems typically exhibit outbursts that last several weeks, after which accretion onto the neutron star ceases and the binary returns to a quiescent state. Usually it remains as such for many years before entering a new active phase. The accretion of matter compresses the neutron star crust, which induces a chain of exothermic electron captures, neutron emissions and density-driven nuclear fusion reactions that take place at several hundreds meters depth \citep{haensel1990a,haensel2008,gupta07}. The energy released in these reactions locally heats the crust, but is eventually conducted over the entire stellar body and radiated away via neutrinos emitted from the dense core, and via thermal X-ray radiation from the neutron star surface \citep{brown1998,rutledge2002}. During quiescent episodes, the thermal X-ray radiation emerging from the hot neutron star can be detected with sensitive X-ray satellites. 

Prior to the onset of an accretion phase, a neutron star is nearly isothermal and the surface radiation therefore offers a probe of the temperature of the stellar core \citep{colpi2001}. Shortly after an accretion outburst, however, the thermal X-ray emission tracks the temperature of the heated crust \citep{ushomirsky2001,rutledge2002}. As the crust cools and restores thermal equilibrium with the core, the thermal X-ray emission decreases and settles at a quiescent base level set by the core temperature \citep{rutledge2002}. This thermal relaxation depends on the composition of the crust, which determines the conduction of heat towards the surface and the core, and on the details of the nuclear reactions, which set the magnitude and depth of the heat sources \citep{brown08}. 

In the past decade, crust cooling has successfully been monitored for four transiently accreting neutron stars: \ks\ \citep{wijnands2001,cackett2010}, \mxb\ \citep[][]{wijnands2003,cackett2008}, \xte\ \citep[][]{fridriksson2011} and \exo\ \citep[][]{degenaar2010_exo2,diaztrigo2011}. All four belong to a rare sub-group of transient LMXBs, the so-called quasi-persistent sources, which exhibit unusually long accretion episodes extending up to decades. These served as prime targets for crust cooling studies, since the prolonged accretion ensures a substantially heated crust. Comparing the X-ray observations with thermal evolution simulations has provided unique insight into the properties of neutron star crusts \citep{shternin07,brown08}. 

An important conclusion drawn from these studies is that the crust efficiently conducts heat such that the ion lattice must have a highly ordered structure \citep{shternin07,brown08}. This poses problems for explaining the ignition of thermonuclear superbursts observed from some accreting neutron stars: these require a critical crust temperature that may be difficult to achieve when heat is efficiently conducted \citep{cooper2005,cumming06}. As a possible solution, it has been hypothesized that accretion deposits more heat inside the neutron star than is accounted for in current models \citep{brown08,gupta07,horowitz2008}. New observations of neutron star crust cooling can confirm the existence, and constrain the magnitude and location, of such possible additional sources of heat \citep{brown08}. Unfortunately, X-ray binaries with long accretion phases are very rare: only a handful of sources serve as potential targets for future studies \citep{wijnands04_quasip}. 

\begin{figure}
 \begin{center}
\includegraphics[width=8.0cm]{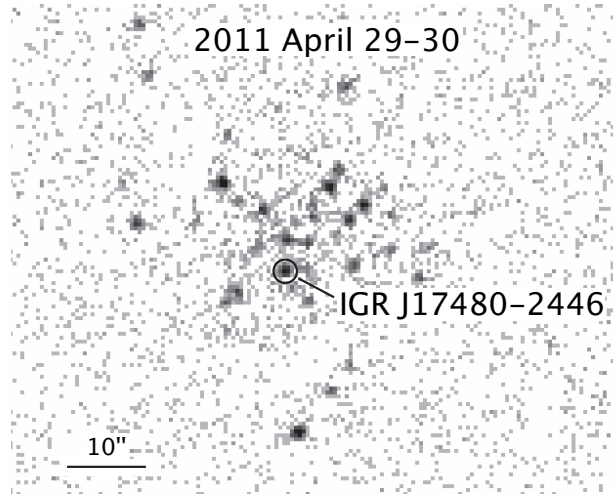}
    \end{center}
\caption[]{{\chan/ACIS image of Terzan 5 obtained on 2011 April 29--30. J1748 is located in the dense cluster core.
}}
 \label{fig:ds9}
\end{figure}

\igr\ (\cxo; J1748 hereafter) is a transient LMXB located in the dense core of the globular cluster Terzan 5 (Fig.~\ref{fig:ds9}). It was initially discovered during \chan\ observations of the cluster performed in 2003 and classified as a low-luminosity X-ray source, possibly a dormant X-ray binary \citep[][]{heinke2006_terzan5}. In 2010 October, the source intensity suddenly increased by orders of magnitude \citep[][]{bordas2010,pooley2010}. The detection of thermonuclear X-ray bursts \citep[][]{chenevez2010_terzan5,linares2011} and coherent X-ray pulsations \citep[][]{strohmayer2010, papitto2010} unambiguously identified J1748 as a neutron star LMXB. Timing analysis of the 11-Hz X-ray pulsations, representing the neutron star's spin period, revealed an orbital period of $\simeq21$~h and a $\simeq 0.4$--$1.5 ~\Msun$ donor star \citep[][]{papitto2010}. 

In 2010 December, $\sim10$~weeks after its onset, the accretion activity ceased and the binary returned to quiescence \citep{deeg_wijn2011_2}. A \chan\ observation obtained about 50 days later, in 2011 February, demonstrated that the quiescent thermal X-ray emission of J1748 was elevated above the base level measured from archival observations taken in 2003 and 2009, by approximately a factor of 4 \citep{deeg_wijn2011_2,deeg_wijn2011}. The associated rise in neutron star temperature (a factor of 1.4), can be interpreted as an accretion-heated crust, although two alternative explanations could be invoked \citep{deeg_wijn2011_2}. First, the neutron star possibly continues to accrete at a very slow rate, causing a variable quiescent flux. Second, the amount of hydrogen and helium left on the neutron star after an accretion phase determine the heat flux that is flowing from the stellar interior towards the surface. For a constant interior temperature, the quiescent thermal emission might therefore differ by a factor of a few due to changes in the envelope composition after an intervening outburst \citep{brown2002}. Both of these scenarios could also account for the enhanced surface temperature observed in 2011.

\section{Observations, analysis and results}\label{sec:results}

\subsection{A new Chandra observation of Terzan 5}\label{subsec:obs}
To further investigate the cause of the elevated thermal X-ray emission detected from J1748 after its 10-week outburst, we obtained a new \chan\ Director's Discretionary Time observation of Terzan 5 on 2011 April 29--30 (obs ID 13252; Fig.~\ref{fig:ds9}). This is $\simeq125$~days into quiescence and $\simeq75$ days since the previous \chan\ observation. The cluster was positioned on the back-illuminated S3 CCD at the nominal target position and the observation was read out in the FAINT timed data mode. We used the \textsc{ciao} software tools (v. 4.3) for the reduction and extraction of data products, following standard analysis procedures. There were no occurrences of background flares during the observation, so that the final net exposure time on our target was 39.5 ks. 

Using the task DMEXTRACT, we extracted source count rates and light curves from a circular region with a radius of $1''$, centered at the best-known position of \igr\ \citep{pooley2010}. Corresponding background events were collected from a circular region with a radius of $40''$, positioned on a source-free part of the CCD located $1.4'$ west of the cluster core. J1748 is detected at a net count rate of $(3.7\pm0.3)\times10^{-3}$~counts~s$^{-1}$ and a total of 146 source photons were collected during the observation. 
Source and background X-ray spectra were obtained using the meta-task SPECEXTRACT, which also generates the appropriate ancillary resposonse files (arf) and redistribution matrix files (rmf). Using the \textsc{ftool} GRPPHA, the spectral data was grouped into bins with a minimum number of 20 photons. The resulting background-corrected spectrum was modeled in the energy range of 0.5--10.0 keV using the software package XSPEC \citep[v. 12.6;][]{xspec}.

\subsection{Quiescent spectra of \igr}\label{subsec:spec}
\begin{figure}
 \begin{center}
\includegraphics[width=8.0cm]{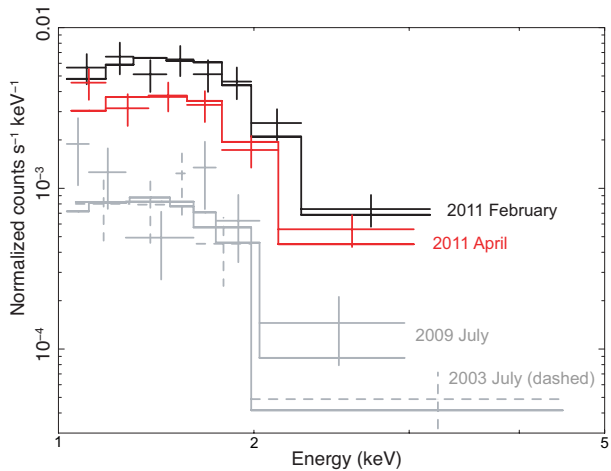}
    \end{center}
\caption[]{{ \chan/ACIS spectra of \igr\ at four epochs. The 2011 data was obtained within a few months after the end of the 2010 October--December outburst and the 2009 data $1.2$~yr prior to the accretion activity. The solid lines indicate best-fits to the neutron star atmosphere model NSATMOS.
}}
 \label{fig:spec}
\end{figure}

Similar to previous observations \citep[][]{deeg_wijn2011_2,deeg_wijn2011}, the X-ray spectrum of J1748 can be adequately fit by an absorbed thermal emission model and does not require the addition of any hard, non-thermal component (Fig.~\ref{fig:spec}). Within XSPEC, we use the neutron star atmosphere model NSATMOS \citep{heinke2006} to fit the thermal emission from the neutron star surface. In this model, we fix the mass and radius of the neutron star to $M=1.4~\mathrm{M}_{\odot}$ and $R=10$~km, respectively, whereas the source distance is set to the current best estimate for Terzan 5: $D=5.5$~kpc \citep{ortolani2007}. The normalization was frozen at the recommended value of 1, which implies that the radiation is emerging from the entire stellar surface. Our only free parameter for the NSATMOS model is thus the neutron star temperature. To take into account the hydrogen absorption along the line of sight, $N_H$, we include the PHABS model.

To chart the evolution in thermal emission and neutron star temperature, we fit the spectrum of this new observation together with \chan\ data obtained on 2003 July 13--14, 2009 July 15--16 and 2011 February 17. These had the same setup as the new observation and were treated following the same reduction and analysis steps used here \citep[for details, see][]{deeg_wijn2011_2,deeg_wijn2011}. Following previous work, we treat the 2003 and 2009 data as one single spectrum \citep{deeg_wijn2011}. We tied the hydrogen column density between the data sets, i.e. this fit parameter is required to be the same amongst the observations, whereas the neutron star temperature could vary freely for the different spectra. This approach resulted in a joint fit value of $N_H=(2.0\pm0.3)\times10^{22}~\nh$ and a reduced chi-square of $\chi_{\nu}^2=0.77$ for 21 degrees of freedom. Fig.~\ref{fig:spec} displays the X-ray spectra observed at the four epochs.

For each data set we deduce the neutron star temperature as seen by an observer at infinity, $kT^{\infty}$, and calculate the total unabsorbed model flux in the 0.5--10 keV energy range, $F_{\mathrm{unabs}}$. The model fits are then extrapolated to the  0.01--100 keV energy range to obtain an estimate of the total emitted bolometric flux, which is translated into a luminosity, $L_{\mathrm{bol}}$, by assuming that the source is located at a distance of $D=5.5$~kpc \citep{ortolani2007}. The results of our spectral analysis are summarized in Table~\ref{tab:spec}, where all quoted errors refer to $90\%$ confidence levels. The evolution of the neutron star temperature as inferred from the spectral fits is plotted in Fig.~\ref{fig:lc}.

\vspace{-0.3cm}
\section{Discussion}\label{sec:discuss}
We analysed a new \chan\ observation of Terzan 5, obtained $\simeq125$~days after the end of the accretion outburst of J1748. The quiescent X-ray spectrum of the source is soft and can be adequately fit by a neutron star atmosphere model NSATMOS with a temperature of $kT^{\infty}=92.1\pm3.1$~eV. This is nearly $10\%$ lower than measured 75 days earlier in 2011 February ($kT^{\infty}=100.9\pm3.3$~eV), yet over $20\%$ higher than the quiescent base level of the source detected in 2003 and 2009 ($kT^{\infty}=72.0\pm3.3$~eV). The thermal X-ray emission and inferred neutron star temperature have thus decreased after the initial enhancement (Fig.~\ref{fig:lc}).

To explain the elevated thermal emission observed in 2011 in terms of a different envelope composition left after the 2010 accretion phase, the neutron star temperature inferred from our new observation should have been similar to the previous measurement, since the composition of the layer would not change in absence of accretion \citep[][]{brown2002}. This is not consistent with our results. Continued accretion is thought to reveal itself through stochastic variability in the X-ray emission on both long (months/years) and short (seconds/days) time scales, and/or via the presence of a hard, non-thermal emission component in the X-ray spectrum \citep{rutledge2001,cackett2010_cenx4,fridriksson2011}. We find none of such features in the \chan\ data of J1748, so there are no obvious indications that the neutron star is accreting at a low level.\footnote{We note that if residual accretion occurs in quiescence, variations in the thermal emission could be due by a changing emission area (a hotspot at the magnetic pole) and may not necessarily imply a changing surface temperature. Given the data statistics, it would not be possible to make this distinction.}

On the other hand, the observed decrease in thermal X-ray emission and neutron star temperature is very reminiscent of the crust cooling observed from the four X-ray binaries with long accretion phases. The data points of J1748 (Fig.~\ref{fig:lc}) are consistent with an exponential decay with an e-folding time of $\sim100-200$~days or a power law with a decay index of order $\sim0.1$. This is similar to the results obtained for the four quasi-persistent LMXBs \citep{cackett2008,cackett2010,degenaar2010_exo2,fridriksson2011}. We therefore consider this the most likely interpretation of the observations. Thus, we propose that the crust of the neutron star in J1748 was heated during its recent 10-week accretion phase and is currently cooling, supporting previous speculations \citep{deeg_wijn2011_2,deeg_wijn2011}.

\subsection{Crust cooling simulations of \igr}\label{subsec:sim}

\begin{figure}
 \begin{center}
\includegraphics[width=8.0cm]{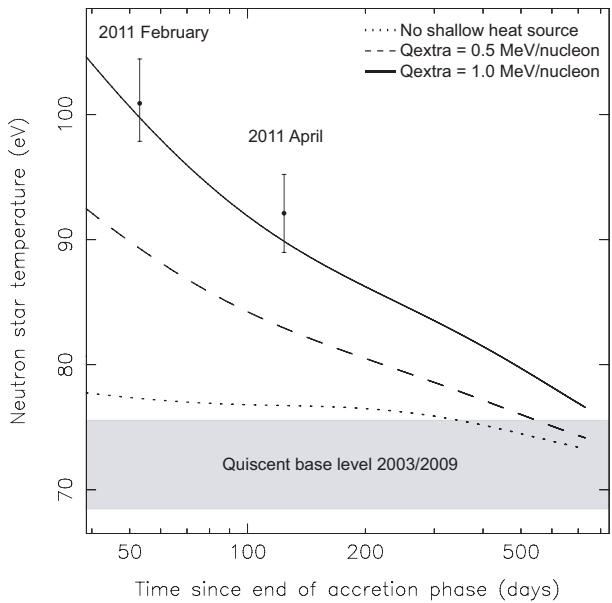}
    \end{center}
\caption[]{{
Evolution of the neutron star temperature of \igr, as inferred from analysis of \chan\ spectral data. The grey shaded area marks the quiescent base level observed prior to the accretion activity of 2010 October--December. The three curves indicate neutron star thermal evolution simulations for different amounts of heat production in shallow crustal layers. 
}}
\label{fig:lc}
\end{figure}

We employed the thermal evolution code of \citet{brown08}, using all the basic physics ingredients and modeling approach described in that work, to simulate neutron star crust cooling curves for J1748 (Fig.~\ref{fig:lc}). We assumed that the outburst had a duration of $t_{\mathrm{ob}}=0.17$~yr and that mass was accreted at a rate of $\dot{M}=2.0\times10^{17}$~g~s$^{-1}$, as inferred from X-ray observations \citep{deeg_wijn2011_2}. Furthermore, the temperature of the neutron star core was fixed at $T=7.0\times10^{7}$~K, which matches the inferred value of the quiescent base level \citep{deeg_wijn2011}. 

Using standard physics input, the crust temperature is too low and the slope of the cooling curve too flat to match the observations (dotted curve in Fig.~\ref{fig:lc}). Motivated by recent theoretical conjectures (see Sec.~\ref{sec:intro}), we also performed simulations with a heat source placed at shallow depth inside the neutron star and varied its strength, $Q_{\mathrm{extra}}$. We position this extra heat at a column depth of $P/g=4.5\times10^{11}$~g~cm$^{-2}$, which roughly corresponds to a density $\rho \simeq 4\times10^{8}$~g~cm$^{-3}$ or about 10~m inside the neutron star \citep{brown08}. This is roughly the depth at which superbursts are thought to ignite. As can be seen in Fig.~\ref{fig:lc}, adding extra heat raises the temperature and steepens the slope of the crust cooling curve (dashed and solid lines). 

For the assumed $\dot{M}\simeq2\times10^{17}~\mathrm{g~s}^{-1}$ \citep{deeg_wijn2011}, we find that a heat source of $Q_{\mathrm{extra}}\simeq1.0$~MeV per accreted nucleon can match the observational data of J1748 (solid curve in Fig.~\ref{fig:lc}). This is substantial compared to the total heat release of $\simeq2.0$~MeV per accreted nucleon that is accounted for by current nuclear heating models \citep[][]{gupta07,haensel2008}. However, the input value of the mass-accretion rate has a notable effect on the resulting crust cooling curve and hence on the strength of the inferred additional heat source. If we increase $\dot{M}$ by a factor of two in our model calculations, which is not implausible given the uncertainties in determining this parameter from X-ray observations \citep{zand07}, the magnitude of the required extra heat decreases to $Q_{\mathrm{extra}}\simeq 0.5$~MeV per accreted nucleon. Nevertheless, even if J1748 accreted at the Eddington limit, additional heat is required to explain the observations as cooling of the accretion-heated crust.

As discussed by \citet{brown08}, heat is primarily conducted in the neutron star crust by electrons, and hence the thermal conductivity is determined by the rate of scatterings between electrons and ions, and between electrons and lattice impurities. The level of impurities in the crust is parametrized by the distribution of ion charges. For the calculations presented in Fig.~\ref{fig:lc}, this was set to $\langle Z^2\rangle - \langle Z\rangle^2=5$, which is consistent with the values inferred from modeling the crust cooling curves of two other neutron stars \citep{brown08} and molecular dynamics simulations \citep{horowitz2009_cond}. We explored simulations with different levels of impurities, but found that this does not affect the resulting crust cooling curve of J1748. This is because, unlike systems with accretion episodes that last for years and heat the entire crust, for short outbursts the heating primarily occurs at low densities, where the heat conduction is controlled by electron-ion scattering. 

We calculated crust cooling curves varying the depth of the extra heat source. We found that the exact depth primarily affects the cooling rate at early times, within a few tens of days after the end of the accretion outburst. J1748 could not be observed with \chan\ on such a time scale, because the source was too close to the Sun at that time. Thus, the location of the extra heat source does not affect the crust cooling curve at the time probed by the observations of J1748. This also implies that our \chan\ observations cannot firmly constrain the depth of the additional heat release, although we can obtain some limits. 

Assuming that the crust cooling curve of J1748 indeed follows a power law decay with an index of $-0.1$ (see above), we estimate that there is an inward-directed heat flux in the crust of $F\simeq3\times10^{21}~\mathrm{erg~cm}^{-2}~\mathrm{s}^{-1}$ \citep[][eq. 12]{brown08}. At any given depth inside the neutron star, it takes a certain characteristic time for heat to diffuse to the surface. Therefore, at a given time there will be a certain depth at which the layer above has already thermally relaxed, while deeper layers are still hot and have not yet started to cool \citep{brown08}. The first \chan\ observation of J1748 was obtained $\simeq50$~days after the end of the accretion outburst: this is the characteristic thermal time scale for layers located at a column depth of roughly $P/g\simeq10^{14}~\mathrm{g~cm}^{-2}$, which corresponds to a matter density of $\rho \simeq 3\times10^{10}~\mathrm{g~cm}^{-3}$ or a depth of $\simeq150$~m \citep{brown08}. Since the heat flux is directed inwards, there must thus be a source of heat exterior to the depth corresponding to the thermal time scale of $50$~days. This implies that there is a heat deposit at less than $\simeq150$~m inside the neutron star \citep{brown08}. This is much shallower than the depth at which most crust nuclear reactions are thought to occur \citep{gupta07,haensel2008,horowitz2008}.

\begin{table}
\caption{Results from fitting the \chan\ spectral data.}
\begin{threeparttable}
\begin{tabular}{c c c c c c c}
\hline \hline
Date & $kT^{\infty}$ & $F_{\mathrm{unabs}}$ & $L_{\mathrm{bol}}$\\
\hline
2003-07-13/14 & \multirow{2}{*}{$72.0\pm3.3$} & \multirow{2}{*}{$1.64\pm0.30$} & \multirow{2}{*}{$0.59\pm0.11$} \\ 
2009-07-14/15 &  &  \\ 
2011-02-17 & $100.9\pm3.3$ & $6.33\pm0.83$ & $2.29\pm0.30$ \\  
2011-04-29/30 & $92.1\pm3.1$ & $4.39\pm0.60$ & $1.59\pm0.21$ \\ 
\hline
\end{tabular}
\label{tab:spec}
\begin{tablenotes}
\item[]Note. $kT^{\infty}$ is given in eV, $F_{\mathrm{unabs}}$ in $10^{-13}~\flux$ (0.5--10 keV) and $L_{\mathrm{bol}}$ in $10^{33}~\lum$ (0.01--100 keV).
\end{tablenotes}
\end{threeparttable}
\end{table}

\subsection{Repercussions}\label{subsec:reper}
The X-ray observations of J1748 obtained after its recent 10-week accretion outburst can be explained as cooling of the accretion-heated neutron star crust. However, regardless of the exact value of $\dot{M}$, this requires a substantial source of heat located at shallow depth inside the neutron star that is not accounted for by existing nuclear physics calculations \citep{gupta07,haensel2008,horowitz2008}. A shallow heat source has also been proven necessary to explain the crust cooling curve of at least one of the quasi-persistent LMXBs \citep[\mxb;][]{brown08}. The presence of substantial heat sources in the outer crustal layers is important for the ignition of superbursts: the extra heat release raises the crust temperature and can thus compensate for the effect of a high thermal conductivity \citep{gupta07,brown08}. 

Our work on J1748 suggests that the crust of a neutron star can become substantially heated during an accretion episode of only a few weeks, and that it is technically feasible to detect the subsequent crust cooling with existing, sensitive X-ray instruments. LMXBs in which the neutron star accretes for weeks or months are much more common than those that undergo extended accretion episodes of years. Future crust cooling studies can therefore draw from a larger sample of transiently accreting neutron stars. The most promising targets are neutron stars that have relatively low core temperatures, 
but exhibit bright accretion episodes so that a substantial amount of matter is consumed and a considerable temperature gradient can develop in the crust \citep{brown1998}. Transient Be/X-ray binaries, which harbor strongly magnetized neutron stars feeding of the matter expelled by their massive Be companions, can potentially also be included in future studies \citep{brown1998}.

Studying a larger sample of crust cooling curves allows to investigate whether the presence of strong, shallow heat sources is a common feature amongst transiently accreting neutron stars. In particular, observations performed within $\sim1$~month after the end of an accretion outburst can further constrain the magnitude and depth of such heat sources \citep{brown08}. The next challenge is then to identify the nature of this extra heat release.

Our crust cooling simulations suggest that \igr\ continues to cool within the next year. New \chan\ observations are planned to confirm this and to put further constraints on the shape of the crust cooling curve.

~\\
\noindent {\bf Acknowledgements.}
The authors are grateful to Harvey Tananbaum and the \chan\ science team for making this observation possible. RW acknowledges support from a European Research Council (ERC) starting grant. Support for EFB was provided by NASA through \chan\ Award Number TM0-11004X. This work was supported by the Netherlands Research School for Astronomy (NOVA). ND and EFB acknowledge the hospitality of the International Space Science Institute (ISSI) in Bern, Switzerland, where part of this work was carried out.

\label{lastpage}
\end{document}